\documentclass[preprint]{aastex}

\newcommand{\subsun}{\mbox{$_{\odot}$}}
\newcommand{\teff}{$T_{eff}$}
\newcommand{\grav}{log($g$)}
\newcommand{\etal}{{\it et al.\/}}

\newcommand{\cband}{C$_2$}

\begin{document}

\title{The Frequency of Carbon Stars Among Extremely Metal
Poor Stars\altaffilmark{1}}

\author{Judith G. Cohen\altaffilmark{2}, 
Stephen Shectman\altaffilmark{3} 
Ian Thompson\altaffilmark{3},
Andrew McWilliam\altaffilmark{3},  
Norbert Christlieb\altaffilmark{4},
Jorge Melendez\altaffilmark{2},
Franz-Josef Zickgraf\altaffilmark{4},
Solange Ram\'{\i}rez\altaffilmark{5}
\& Amber Swenson\altaffilmark{2} }

\altaffiltext{1}{Based in part on observations obtained at the
W.M. Keck Observatory, which is operated jointly by the California 
Institute of Technology, the University of California, and the
National Aeronautics and Space Administration.}

\altaffiltext{2}{Palomar Observatory, Mail Stop 105-24,
California Institute of Technology, Pasadena, Ca., 91125, 
jlc(jorge)@astro.caltech.edu, aswenson@its.caltech.edu}

\altaffiltext{3}{Carnegie Observatories, 813 Santa
Barbara Street, Pasadena, Ca. 91101, andy(ian,shec)@ociw.edu)}

\altaffiltext{4}{Hamburger Sternwarte, Universit\"at
Hamburg, Gojenbergsweg 112, D-21029 Hamburg, Germany,
fzickgraf@hs.uni-hamburg.de,
nchristlieb@hs.uni-hamburg.de}

\altaffiltext{5}{Spitzer Science Center, Mail Stop 100-22,
California Institute of Technology, Pasadena, Ca., 91125,
solange@ipac.caltech.edu}

\begin{abstract}

We demonstrate that there are systematic scale errors in
the [Fe/H] values determined by the Hamburg/ESO Survey
(and by inference by the HK Survey in the past) for certain extremely metal
poor highly C-enhanced giants.
The consequences of these scale errors are that
a) the fraction of carbon stars at extremely
low metallicities has been overestimated
in several papers in the recent literature
b) the number of extremely metal poor stars known is
somewhat lower than has been quoted in the recent
literature c) the yield for extremely metal poor stars
by the HES Survey is somewhat lower
than is stated in the recent literature.  A preliminary
estimate for the frequency of Carbon stars among the giants in the HES
sample with $-4 <$ [Fe/H] $ < -2.0$ dex is 7.4$\pm2.9$\%; adding an
estimate for
the C-enhanced giants with [C/Fe] $> 1.0$ dex without detectable 
\cband\ bands raises the fraction to 14$\pm4$\%.

We rely on the results of an extensive set of
homogeneous detailed abundance analyses of stars expected to have
[Fe/H] $\le -3.0$ dex  selected from the HES
to establish these claims.
We have found that 
the Fe-metallicity of the cooler (\teff $\lesssim 5200$~K) C-stars as derived
from spectra taken with HIRES at Keck are a factor of $\sim$10
higher than those obtained via the algorithm used
by the HES project  to analyze the
moderate resolution follow-up spectra, which is
identical to that used until very recently by the HK Survey. 
This error in Fe-abundance estimate for C-stars 
arises from a lowering of the emitted flux
in the  continuum bandpasses of the KP 
(3933~\AA\ line of Ca~II) and particularly the HP2 (H$\delta$) indices
used to estimate [Fe/H] 
due to absorption from strong molecular bands.

\end{abstract}

\keywords{Galaxy:halo, stars:carbon, Galaxy:abundances}

\section{Introduction}

We are engaged in a large scale project to find additional 
extremely metal poor stars in the halo of our galaxy.
The major existing survey for very metal-poor stars is the HK survey
described in detail by Beers, Preston \& Shectman (1985, 1992). 
The stellar inventory of this survey has been scrutinized with
considerable care over the past decade, 
but, as summarized by \cite{beers98}, only roughly 100
are believed to be extremely metal poor (henceforth EMP), 
with [Fe/H] $\le -3.0$ dex\footnote{The 
standard nomenclature is adopted; the abundance of
element $X$ is given by $\epsilon(X) = N(X)/N(H)$ on a scale where
$N(H) = 10^{12}$ H atoms.  Then
[X/H] = log$_{10}$[N(X)/N(H)] $-$ log$_{10}$[N(X)/N(H)]\subsun, and similarly
for [X/Fe].}.
We are therefore exploiting the database of the
Hamburg/ESO Survey (HES) for this purpose.   The HES is an 
objective prism survey from which it is
possible to efficiently select EMP stars \citep{christlieb03}.
The existence of a new
list of candidates for EMP stars 
with [Fe/H] $< -3$ dex 
selected in an automated and unbiased manner
from the HES, coupled with
the very large collection area and efficient high resolution echelle
spectrographs of the new generation of large telescopes,
offers the possibility for a large
increase in the number of EMP stars known
and in our understanding of their properties.

We have obtained and analyzed spectra with HIRES \citep{vogt94}
at the Keck I Telescope of a large number of EMP
candidates selected from the HES.
The normal procedures outlined by \cite{christlieb03}
to isolate EMP stars from the
candidate lists produced by the HES were followed. In an effort
to avoid selection biases, these differ from the criteria
adopted by the HK Survey; see \cite{christlieb03} for details.
Candidate EMP stars selected from the HES were
vetted via moderate resolution spectroscopy
at large telescopes
to eliminate the numerous higher abundance interlopers.
Most of the follow up spectra for the stars discussed here were obtained
with the Double Spectrograph \citep{dbsp} on the Hale Telescope
at Palomar Mountain, denoted P200, (a few are from the Boller and Chivens spectrograph on the 
Clay and Baade Telescopes at the Las Campanas Observatory) during the 
period from 2001 to the present.  We intend to observe all candidates to the magnitude
limit of the HES (B $\sim$ 17.5) in our fields; observations are now
complete in $\sim$990 deg$^2$, complete to B=16.5 in 
an additional $\sim$700 deg$^2$,
and approaching completion in the remaining fields.

These follow up spectra are used to determine an accurate measure of
the metallicity of the star, much more so than is possible 
to derive from the low resolution objective prism spectra of the HES itself. 
This is accomplished via a combination of strength of absorption in 
H$\delta$ (determining \teff) and in the 
Ca~II line at 3933~\AA\ (the KP index, which determines
[Fe/H], once \teff\ and hence \grav\ are specified).  A calibration between the
strength of the indices
and metallicity is required, and is generally derived from
literature searches for high resolution abundance studies of relevant stars.
We denote the resulting metallicity
value as [Fe/H](HES).  The specific algorithm adopted by the HES is
described in \cite{beers99} and is essentially identical to that
used by the HK Survey until recently; the latest updates to the algorithm
as used by the HK Survey are described in \cite{rossi05}.

\section{Systematic Calibration Problems in the Metallicity Scale of the
HES and HK Survey}

Stars were chosen for  observation
at high resolution with HIRES
primarily on the basis of low predicted metallicity;
every star with [Fe/H](HES) $\le -2.9$ dex
north of $\delta -25^{\circ}$ was put on the HIRES observing list.
Spectra have now been obtained for more than 55 EMP candidates from the HES.
In all cases the stellar parameters have been determined by JGC from
broad band (V-I, V-J and V-K) photometry and theoretical isochrones
with no reference to the spectra themselves. Insofar as possible,
the procedures, the codes, the model 
atmospheres \citep[we use those of][]{kurucz93} and
the atomic data used to reduce the HIRES echelle spectra
and to carry out the detailed abundance analysis are identical, and
the analyses are thus as homogeneous as possible.  \cite{cohen04} present
full details of the analysis and
results for a large sample of EMP dwarfs from the HES, while
\cite{cohen05b} will present abundance analyses for 13 of the 16 
known carbon stars from this HES sample; two of these have not 
yet been observed with HIRES.   Fifteen 
C-normal giants have been analyzed to date, and these 
will appear in a future publication \citep{cohen05d}.

Our operational definition of a carbon-star (C-star) is one
whose spectrum shows bands of \cband.  
The P200 DBSP spectra mostly extend to 5300~\AA, hence the prominent
\cband\ band at 5160~\AA\ is included. We are reasonably certain, 
as will be discussed
in \cite{cohen05c}, through inspection of the regions
of both \cband\ and CH in these followup spectra,
that there are no additional giant C-stars in the Palomar sample.  
If no \cband\ bands are detected, but [C/Fe]$ > 1$ dex, we denote a
star to be C-enhanced.
Also we denote stars with \teff\ $ > 6000$~K as ``dwarfs'', while all cooler
stars are called ``giants''.

The difference between the [Fe/H] derived from analysis of the HIRES spectrum versus
that obtained by applying the algorithm of \cite{beers99} to the moderate
resolution follow up spectra for the set of 497 candidate EMP stars with
moderate resolution follow up P200 spectra is shown in
Fig.~\ref{fig_delta_feh}.
The giants with normal C and the warmer C-giants show
good agreement; [Fe/H](HES) inferred from the
moderate resolution spectra is a reliable indicator of the  Fe-metallicity
found from analysis of HIRES spectra.
Getting the dwarfs correct is harder as metal lines become weaker
at their higher \teff.  Also we have adopted a \teff\ scale for them
which is hotter than that used in most earlier analyses which provided the
calibration of the \cite{beers99} relation for EMP dwarfs
\citep*[see, e.g.][]{norris96,ryan96}.
In the mean, [Fe/H](HES) systematically underestimates our HIRES-based 
Fe-metallicity by 0.37 dex for EMP dwarfs, corresponding to a systematic 
difference in adopted \teff\ of $\sim$400~K.

\begin{figure}
\epsscale{1.0}
\figurenum{2}
\plotone{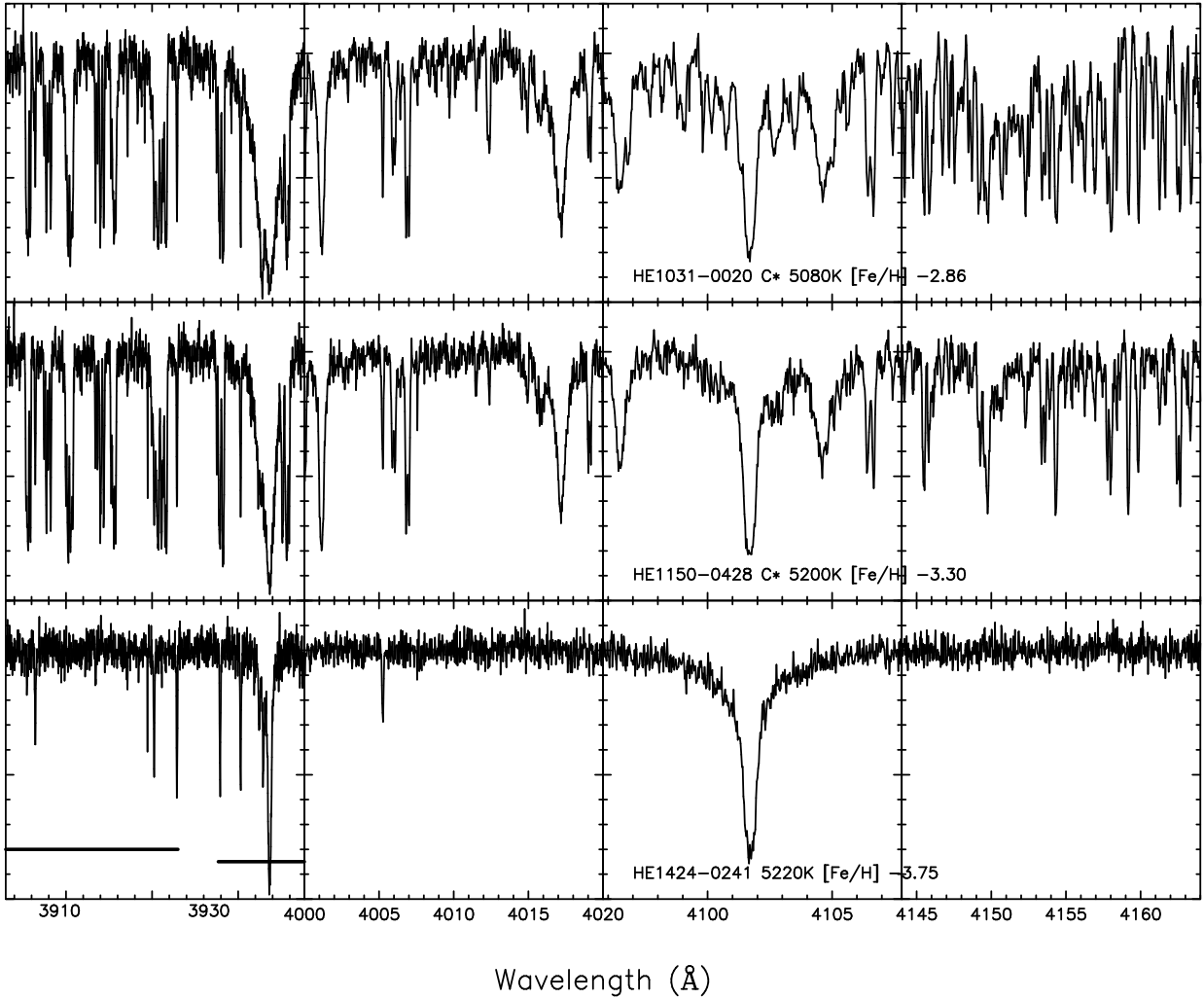}
\caption[]{Four spectral regions from the HIRES spectra of two C-stars 
(two upper rows) and an C-normal EMP giant.  Each star has
\teff $\sim 5150$~K and each has [Fe/H](HES) $< -3.2$ dex.
The HIRES derived [Fe/H] are shown in the label for each star.  
The left column shows the region of the
blue continuum and the feature bandpass for the
KP index as defined in \cite{beers99}; these are indicated by the horizontal
lines in the bottom left panel.  The next column 
is the red continuum bandpass for the KP
index, which is also the blue continuum bandpass for the HP2 index.  The
next column shows the feature bandpass for the HP2 index measuring the
strength of H$\delta$.  The
right column is the red continuum bandpass for the HP2 index, which is heavily
contaminated by molecular features in the C-stars. The feature bandpasses
are taken as 12~\AA\ wide. The vertical scale is
identical for each panel, 0.0 to 1.2 for the normalized flux.
\label{fig_spec}}
\end{figure}

However, Fig.~\ref{fig_delta_feh} shows that
for the cooler C-giants (\teff\ $\lesssim 5200$~K),
[Fe/H](HES) substantially underestimates  our HIRES-based
Fe-metallicity by $\sim$1 dex.  \cite{preston01} suggested the
presence of a systematic error of comparable size (up to 1 dex) for the
sample of C-stars they analyzed from the HK Survey.  This is
a very large systematic error, much too large to be caused
by problems in the \teff\ scale, and so we attempt to understand
what might be causing it.
In Fig.~\ref{fig_spec} we show sections of the 
HIRES spectra of three EMP candidates
from the HES shifted into the rest frame.
These stars all have \teff $\sim 5150$~K. Two are C-stars, the 
third is a genuine EMP giant with weak CH.  Each of these stars has 
[Fe/H](HES) $< -3.2$ dex.
The red and blue continuum and the feature bandpass are shown for the
KP and the HP2 indices used to determine [Fe/H](HES),
with a feature bandpass 12~\AA\ wide for each; see
\cite{beers99} for details of the index definitions.  The figure
clearly shows the source
of the problem afflicting the C-stars -- the ``continuum''
bands are full of strong molecular absorption, particularly the
red continuum band for the HP2 index.
If the HP2 index is underestimated because the continuum is depressed,
then the star is assumed to be cooler than it actually is, and the
resulting [Fe/H](HES) for a fixed KP index will be too low.  Furthermore,
the blue continuum region of the KP index also shows strong CH absorption (the
big chunks missing from the spectra of the C-stars in the left column blueward
of the 3933~\AA~CaII line), and hence the
abundance indicator KP will also be underestimated.  The derived
[Fe/H](HES) obtained using the calibration of \cite{beers99} 
will thus be substantially reduced below its true value for such C-stars.  Because
the absorption in the relevant spectral regions arises from both CN and CH,
the magnitude of this effect depends on additional factors 
such as the C-enhancement and the C/N ratio as well as on \teff.

Several tests have been performed to verify this.  First
we checked that the measured KP and HP2 indices
for C-stars (and for C-normal stars) can be reproduced
to within their uncertainties from the
much more precise HIRES spectra.  
We also checked that adding back the missing flux
removed from the continuum by absorption in the 
sidebands significantly increases KP and particularly the
HP2 indices above the
measured values, by factors of two or more.  Finally we checked
that by so altering the KP and HP2 indices we derive a
significantly higher value for
[Fe/H](HES) which is much closer to the that obtained by the detailed
abundance analyses for the cooler C-stars.

\section{Discussion and Implications}

\begin{figure}
\epsscale{0.9}
\figurenum{1}
\plotone{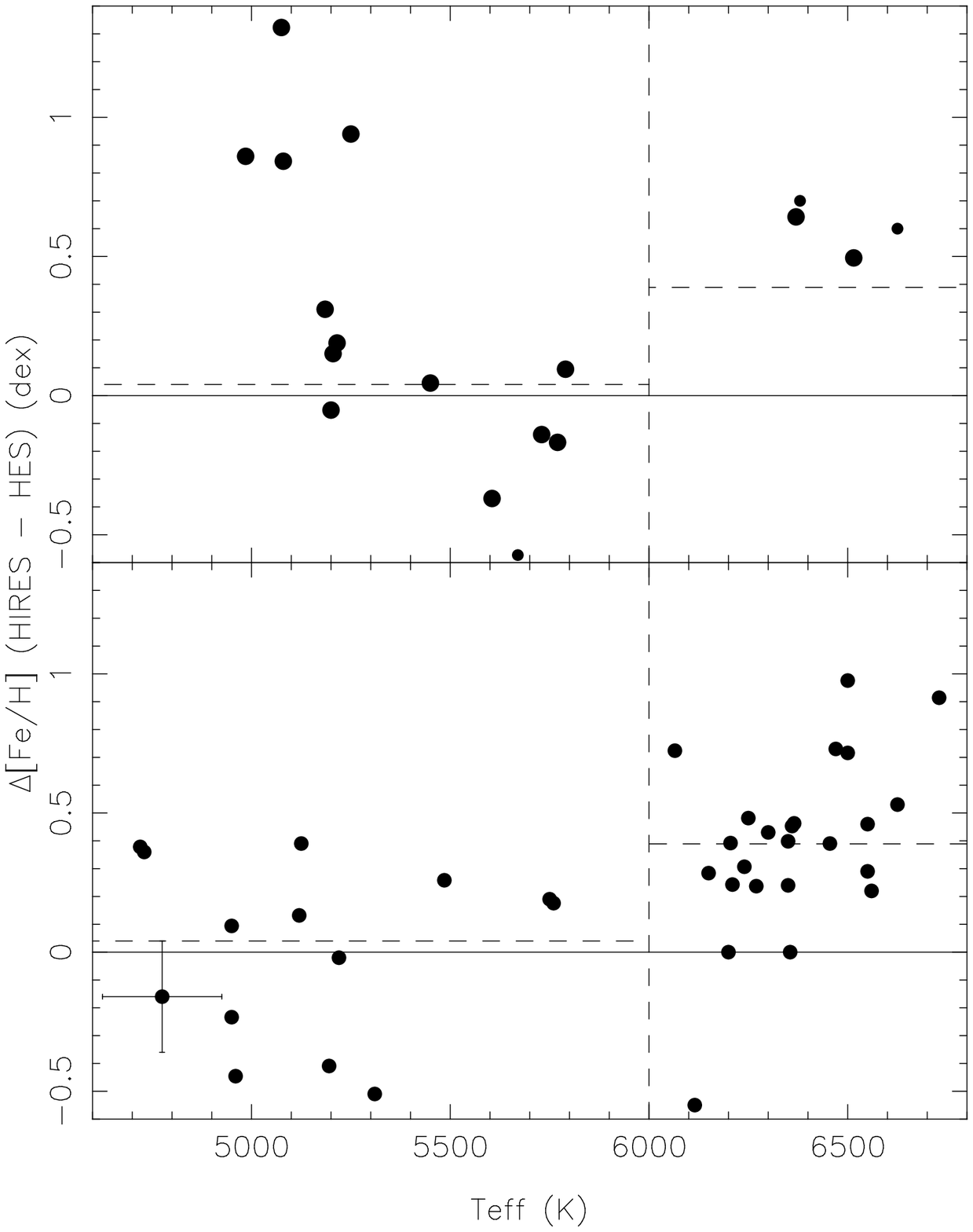}
\caption[]{The difference between [Fe/H](HES) and [Fe/H](HIRES) is shown
as a function of \teff\ for the C-stars (upper panel)
(the smaller symbols denote C-enhanced stars) and for the C-normal
stars (lower panel) for those EMP candidates from the HES with analyses based
on Keck/HIRES spectra.  The vertical dashed line separates the giants
from the dwarfs, while  the horizontal dashed lines are represent
the mean $\Delta$ for the C-normal giants and for the C-normal
dwarfs.  A typical error for a C-normal giant is shown.
\label{fig_delta_feh}}
\end{figure}

\begin{figure}
\epsscale{0.8}
\figurenum{3}
\plotone{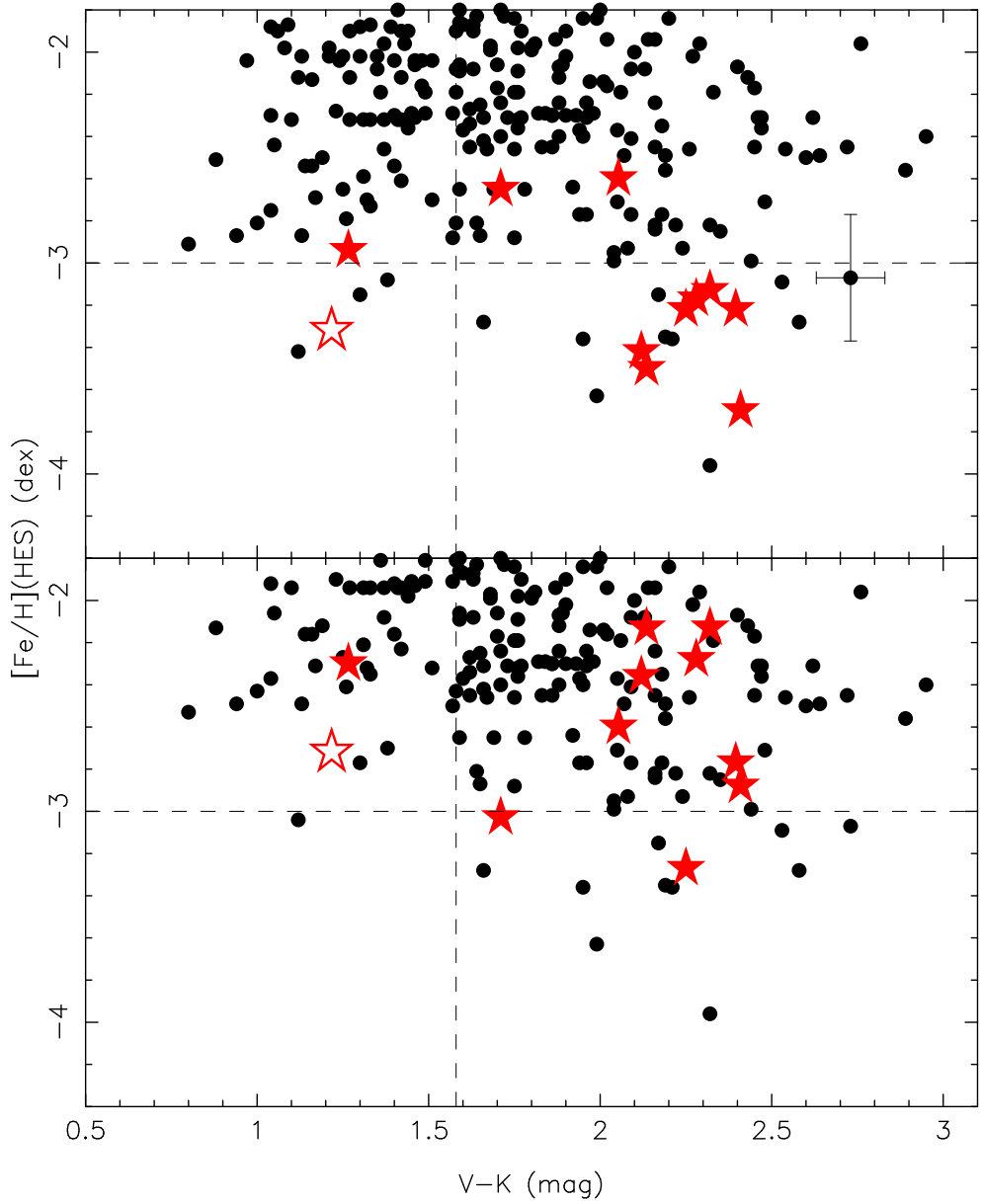}
\caption[]{
A plot of [Fe/H](HES) versus V-K for a sample of 489 EMP
candidates from the HES with moderate resolution
spectra from the Double Spectrograph at the Hale Telescope
(filled circles, limited to stars with [Fe/H](HES) $< -1.8$ dex). 
The C-stars
are from this sample are indicated by filled stars; the C-enhanced star is 
shown as an open star.  In the upper panel
the C-stars are plotted at their [Fe/H](HES) values, while in the lower 
panel, at their [Fe/H](HIRES) values.
The C-normal dwarfs have also been shifted towards higher
[Fe/H] by 0.38 dex in the lower panel.
A typical error for a EMP giant with normal
C is shown for a single star in the upper panel.
The vertical dashed line separates the giants
from the dwarfs.  The horizontal dashed line indicates the EMP
cutoff at [Fe/H] = $-3.0$ dex.
\label{fig_feh_vmk}}
\end{figure}

An underestimate of a factor of
$\sim$1 dex in the deduced value of [Fe/H](HES) for the cooler C-stars
will have significant effects.
Fig.~\ref{fig_feh_vmk} shows [Fe/H](HES) versus V-K for a sample of 489 EMP
candidates from the HES with moderate resolution
spectra from the Double Spectrograph at the Hale Telescope 
\citep[details will appear in][]{cohen05c}.  
The 10 known C-stars and the 1 C-enhanced star with HIRES 
analyses\footnote{The C-enhanced star is HE0024--2523, see 
\cite{lucatello03} and references therein.} from this sample are indicated.
In the upper panel, these stars are plotted
at their [Fe/H](HES) values, while in the lower panel
they are plotted at their [Fe/H](HIRES) as determined from detailed
abundance analyses\footnote{Two of the C-stars in the P200 sample have not
been observed with HIRES;
they are shown with the appropriate offset determined from
Fig.~\ref{fig_delta_feh}.}.  Although at their
nominal Fe-metallicities the C-stars dominate the population of the
giants below [Fe/H](HES) $-3$ dex, 
using the results from analysis of high resolution spectra in the lower
panel the frequency of C-stars 
among the most metal poor EMP stars is considerably reduced.  

It is our contention that, as shown in Fig.~\ref{fig_feh_vmk},
this underestimate of the metallicity of the cooler
C-stars by the algorithm of \cite{beers99}, used by both the HES
and in the past by the HK Survey,  produces a spurious high 
frequency of C-stars among EMP stars  (see Fig.~\ref{fig_delta_feh}).
Using the [Fe/H](HES)  values
for the C-stars would yield an apparent C-star fraction of 33\% for
[Fe/H] $\le -3$ dex, while using the HIRES Fe-metallicities,
a value a factor of 2.4 smaller is obtained.  
There are 122
HES giants with [Fe/H](HES) $< -2.0$ dex in our sample, 
suggesting a C-star frequency of  7.4$\pm2.9$\% for EMP stars.  
Adding in the fraction of of C-enhanced stars 
among giants with [Fe/H](HES) $< -2.0$ dex
found by \cite{cohen05a} of 6.5$\pm2.7$\%,
one obtains a total fraction of C-rich stars with
[C/Fe] $> 1.0$ dex of 14$\pm4$\% among the giants in our HES EMP sample with 
[Fe/H](HES) $< -2.0$ dex, smaller than the value of 25\% for stars with
[Fe/H]  $< -2.5$ dex given by \cite{marsteller}.
(We have derived this fraction for the C-enhanced stars
among the dwarfs in our sample with [Fe/H](HES) $< -2.0$ dex; it has
the same value, as will
be reported in Cohen et al 2006a.)  We are currently analyzing
larger samples to refine this fraction.
It will probably be necessary to include 
an indicator of the strength of the molecular bands 
\citep[most easily the G band of CH, i.e. the GP index already introduced in][]{bee85}
with the standard KP and HP2 indices in the calibration algorithm
to obtain valid Fe-abundances and
maximum information from the HES and HK Survey samples
and/or to replace the HP2 index with
a combination of V from the HES and J or K from 2MASS; the latter was
not available at the time the HK Survey defined their metallicity 
calibration algorithms, \citep*[see, e.g.][]{rossi05}.

The metallicity distribution function is very sharply declining
among halo stars at the lowest Fe-metallicities.
Thus the systematic errors we have found in the calibration of the HES and
by inference the HK metallicity scale of \cite{beers99}, at least
until quite recently \citep*[see][]{rossi05},
will also lead directly to systematic
overestimates of the number of EMP stars and of the yield
for EMP stars by these two major surveys.  
We are currently
evaluating in detail the impact of these calibration errors
on such issues.

\acknowledgements

The entire Keck/HIRES user community owes a huge debt
to the many other people who have
worked to make the Keck Telescope and HIRES a reality and to
operate and maintain the Keck Observatory.  
JGC and JM are grateful for partial support from  NSF grant
AST-0205951.  JGC is grateful for support from
the Ernest Fullam Award of the Dudley Observatory which helped 
initiate this work.
The work of N.C. and FJZ is supported by Deutsche
   Forschungsgemeinschaft (grants Ch~214-3 and Re~353/44). N.C.
   acknowledges support through a Henri Chretien International
   Research Grant administered by the American Astronomical
   Society.

{}

\clearpage

\end{document}